\documentclass[12pt]{article}
\usepackage{latexsym}

\newcommand{\be}{\begin{equation}}
\newcommand{\ee}{\end{equation}}

\begin{document}

\title{Quark-composites approach to QCD}

\author{\\
  Sergio Caracciolo            \\
  {\it Scuola Normale Superiore} \\
  {\it and INFN -- Sezione di Pisa}  \\
  {\it I-56100 Pisa, ITALIA}          \\ 
 \and \\
   Gianni De Franceschi            \\
  {\it Universit\`a di Roma I  La Sapienza }\\
  {\it and INFN -- Sezione di  Frascati}  \\
  {\it Piazzale Aldo Moro 2, I-00185 Roma, ITALIA}          \\
 \and \\
   Fabrizio Palumbo         \\
  {\it INFN -- Laboratori Nazionali di Frascati}  \\
  {\it P.~O.~Box 13, I-00044 Frascati, ITALIA}          
   }

\date{}

\maketitle

\begin{abstract}
\noindent
We present a perturbative approach to QCD based on quark
  composites, that is barions and mesons, as fundamental variables.
\end{abstract}

\clearpage

We present an approach to QCD where quark composites with hadronic
 quantum numbers are assumed
as integration variables in the Berezin integral which defines the
 partition function~\cite{Palu}.  The main motivation is to overcome
 the present impasse whereby the confinement of quarks and the calculation
of hadronic processes are done in different, incompatible frameworks. 

Such an approach is made possible by two results. The first is that
for some   trilinear quark composites, and specifically for the
nucleons,  the integral obtained by the change of 
variables is the Berezin integral over the composites themselves. The
free action of
these composites is the Dirac action.  The second result
concerns the free action of mesonic composites, whose form can be
found on the basis of the analogy with simpler models (which are
potentially interesting for the physics of condensed matter), even
though the integral over these variables is very complicated. Since
the free actions of all these hadronic composites 
are irrelevant operators, they can be freely added to the standard QCD action,
thus providing  the basis for a perturbative expansion, which to lowest order yields
the one pion exchange nucleon-nucleon interaction and the pion-pion interaction of
the chiral models, while the quarks do not propagate.

 One way to perform the change of variables is to assume as new variables the nucleons
at the sites of a sublattice of spacing $2a$, and the chiral mesons at the sites of the
staggered sublattice. We have no space to present such a unified treatment, and then we
will illustrate the two cases
separately, starting from the nucleons.
Consider the QCD partition function
\be
Z= \int [dU][d\overline{\lambda}\,d\lambda] \exp(-S_G -S_q)
\ee
where $S_G$ is the pure gluon action and $S_q$ is the action of the
quarks interacting with the gluons
\be
S_q(r,m_q,U)= a^4\sum_x \bar{\lambda}(x) Q_{x,y}(r,m_q,U)\lambda(y).
\ee
In the last equation the quark wave operator $Q$ is given by
\be
Q_{x,y}(r,m_q,U)=-{ 1\over 2a}\sum_{\mu}(r-\gamma_{\mu})U_{\mu}(x)\delta_{y,x+\mu}
+\left(m_q+ {4r \over a}\right) \delta_{x,y},
\ee
with $0< r \leq 1$ the Wilson parameter,
$U_{\mu}$ is the link variable associated to the gluon field and 
$m_q$  the quark mass, and we followed the standard conventions.
We want to evaluate the partition function by assuming as new integration variables the nucleon composites~\cite{Ioff}
\begin{equation}
\psi_{\tau\alpha}= -{2\over3}k_N^{1/2}a^3
\delta_{\tau\tau_2}\epsilon_{\tau_1\tau_3} (\gamma_5
\gamma_{\mu})_{\alpha\alpha_1} ({\cal C} \gamma_{\mu})_{\alpha_2\alpha_3}
\epsilon_{a_1a_2a_3} \lambda^{a_1}_{\tau_1\alpha_1} 
\lambda^{a_2}_{\tau_2\alpha_2} \lambda^{a_3}_{\tau_3\alpha_3}.
\end{equation}
In the above equation and in the following the summation over repeated indices is
understood, ${\cal C}$ is the charge conjugation matrix, the $\lambda^a_{\tau \alpha}$ are
the quark fields with color, isospin and Dirac indices $a$,
$\tau$ and $\alpha$ respectively, related to the up and down quarks
according to $\lambda^a_{1\alpha}=u^a_{\alpha},\;\;\;\lambda^a_{2\alpha}=d^a_{\alpha}$.
Correspondingly $\psi_{1\alpha},\psi_{2\alpha}$ are the proton, neutron fields which
 with the above definition transform like the quarks
under isospin, chiral and O(4) transformations.

There are altogether 8 $\psi$. Since monomials of less than 8 $\psi$ cannot obviously contain all the $\lambda$, there can be at most one  monomial $\Psi$ which
contains all of them, and with a given ordering for the product $\Lambda$ of all the quark field components at a given site we have
\be
\Psi= \psi_{1,1}... \psi_{1,4}\psi_{2,1}...\psi_{2,4}= J \Lambda,\;\;\;J= k_N^4\;a^{24}\;2^{22}
\cdot 3^3 \cdot5.
\ee
Now $\Lambda$ is the only monomial with nonvanishing Berezin integral
\be
\int   [d\lambda] \Lambda = 1,
\ee
so that, if we assume for the $\psi$ the same rule of integration,
\be
\int   [d\psi] \Psi = 1,
\ee
we have for an arbitrary function $g$
\be
\int [d \lambda] g(\psi(\lambda))= J \int [d \psi] g(\psi).
\ee
One can evaluate more general integrals involving the quarks as well as the nucleon
composites. They vanish unlesss the number of quark fields is equal to  $3 \cdot $integer. An
 example of such an integral is 
\be
\int [d \lambda]  \lambda^{a_1}_{\tau_1 \alpha_1} \lambda^{a_2}_{\tau_2 \alpha_2}\lambda^{a_3}_{\tau_3 \alpha_3}  g(\psi(\lambda)) =
\sum_{\tau \alpha} f^{a_1a_2a_3}_{\tau \alpha \tau_1 \alpha_1 \tau_2 \alpha_2 \tau_3 \alpha_3}  J \int [d \psi]  \psi_{\tau \alpha}g(\psi),
\ee
where  the transformation functions $f$ are
\begin{eqnarray}
f^{a_1a_2a_3}_{\tau \alpha \tau_1 \alpha_1 \tau_2 \alpha_2 \tau_3 \alpha_3}  &=& -{ 1\over 96}
a^{-3}k_N^{-1/2} \epsilon^{a_1a_2a_3} \left[\delta_{\tau \tau_2} \epsilon_{ \tau_1 \tau_3}
(\gamma_5 \gamma_{\mu})_{\alpha_1 \alpha} ( \gamma_{\mu} {\cal C}^{-1})_{ \alpha_2 \alpha_3}\right.
\nonumber\\
 & & \left. + \delta_{\tau \tau_1} \epsilon_{ \tau_2 \tau_3} 
 (\gamma_5 \gamma_{\mu})_{\alpha_2\alpha}
(\gamma_{\mu} {\cal C}^{-1})_{ \alpha_1 \alpha_3} \right]. \label{h}
\end{eqnarray} 
To evaluate the partition function perturbatively, we add to the standard action the Dirac action  $S_N(r_N,m_N)$ constructed for the nucleons in the
same way as for the quarks, and then consider $S_q$ as a perturbation
\be
Z=\prod_x J^2 \int[dU] \exp(-S_G) \int[d\bar{\psi}d\psi]  \sum_{n=0}^{\infty} { 1\over (3n)!}
\left(-S_q\right)^{3n}\;\exp\left( -S_N \right).
\ee 
Since the factor $(S_q)^{3n}$ yields a factor $k_N^{-n}$,
because of the dependence on $k_N$ of the transformation functions,  we have an expansion
in inverse powers of $k_N$. We should emphasize that we do not need  treating the gauge fields
perturbatively.  As an example we report 
the first order contribution to the effective action of the nucleons, which  results to be a pure renormalization
\be 
\left( S_N \right)_1 = - { 3 \over 8}{ 1\over 24^2}  ( 2+ r^2) k_N^{-1}S_N(r'_N,m'_N),\;\;\;
r'_N= r {2 r^2 +1\over 2 +r^2}.
\ee
Notice that $r_N$ satisfies the same restriction as r, and in particular $r'_N=0,1$ for $r=0,1$.

Let us now come to the chiral composites 
\begin{equation} 
  \vec{\pi} = i\,\sqrt{k_{\pi}}\,a^{2}\,\overline{\lambda} \gamma_5 \vec{\tau} \lambda
,\;\;\; \sigma = \sqrt{k_{\pi}}\,a^{2}\,\overline{\lambda} \lambda.
\end{equation}
Their action, invariant under the chiral transformations over the quarks apart
from a linear breaking term, is ~\cite{Cara}
\begin{equation}
S_{C} =   {a^4 \over 2} \sum_{x,y}\left[ \vec{\pi}_x C_{xy} \vec{\pi}_y  +
\sigma_x C_{xy} \sigma_y  \right] - m a^2 \sum_x \sigma_x\label{actionC2} 
\end{equation}
where 
\be
 C = {\rho^4\over a^2} {1\over  a^2  \Box - \rho^2}.
\ee
The inverse of this wave operator is static, but the
propagator of quadratic composites {\it is not the inverse of the wave
operator}.  This is due
to the fact that, even though also the mesonic composites can be
introduced  as integration variables, the resulting integral does not
reduce  in general to a Berezin or to an ordinary integral.  
To overcome the difficulty of calculations with such an integral, auxiliary
bosonic fields $\vec{\chi}$ and $\chi_0$ have been introduced by means
of the Stratonovitch-Hubbard representation. The
integration over the quarks produces an effective action $S_\chi$ for these fields.
A perturbative series is generated by a saddle point method whose
asymptotic parameter is the inverse of the square root of the number
of quarks components $\Omega$, which for up and down quarks is 24.
The minimum of $S_\chi$ is achieved with a breaking of the $O(4)$
invariance, an invariance which follows from the chiral transformations on the
quark-fileds, and, because of the explicit breaking term we introduced, 
a non-zero expectation value is given to $\chi_0$. 
As a result we get
\be
S_{\chi} = S_{\chi}^{free} + \Delta S_{\chi} 
\ee
where
\be
S_\chi^{free} = -{1 \over2} a^4 \sum_{x,y} \vec{\chi}_x
 \left( \Box_{xy} -  m_{\pi}^2 \delta_{xy}  \right)\vec{\chi}_y   
 -{1 \over2} a^4 \sum_{x,y} \theta_x
 \left({ a^2\over 2 \rho^2} \Box_{xy} -   \delta_{xy}  \right)\theta_y 
\ee
with
\be
  m_{\pi}^2 = {m \rho\over a}
\ee
which shows how $m$ must be moved with the lattice spacing $a$ in
order to keep finite the meson mass.
$\theta$ is the fluctuation, up to a scale factor, 
of the $\chi_0$-field around its expectation value and
$\Delta S_{\chi}$ can be given as a renormalizable  expansion in
inverse powers of $\sqrt{\Omega}$, whose terms can be easily expressed
as an expansion in inverse powers of $k_\pi$.

\end{document}